\documentstyle[prl,aps,twocolumn,epsfig]{revtex}
 \newcommand{\mytitle}[1]{
 \twocolumn[\hsize\textwidth\columnwidth\hsize
 \csname@twocolumnfalse\endcsname #1 \vspace{1mm}]}
 \newcommand{\beq}{\begin{equation}}
 \newcommand{\eeq}{\end{equation}}
 \newcommand{\bea}{\begin{eqnarray}}
 \newcommand{\eea}{\end{eqnarray}}
 \newcommand{\pdag}{{\phantom{\dagger}}}
\begin{document}
\draft
\mytitle{
\title{Quantum Phase Transition in a Multi--Level Dot}
\author{Walter Hofstetter$^{1}$ and Herbert Schoeller$^{2}$}
\address{
$^1$Theoretische Physik III, Elektronische Korrelationen und Magnetismus, 
Universit\"at Augsburg, D-86135 Augsburg, Germany\\
$^2$Theoretische Physik A, Technische Hochschule Aachen, D-52056 Aachen,
Germany}
\date{\today}
\maketitle
\begin{abstract}
We discuss electronic transport through a lateral quantum dot 
close to the singlet--triplet degeneracy in the case of a \emph{single} conduction 
channel per lead. By applying the Numerical Renormalization Group, 
we obtain rigorous results for the linear conductance and the 
density of states. 
A new \mbox{\emph{quantum phase transition}} of the Kosterlitz--Thouless type is found, with 
an exponentially small energy scale $T^*$ close to the degeneracy point. 
Below $T^*$, the conductance is strongly suppressed, corresponding to a universal dip in the 
density of states. This explains recent transport measurements. 
\end{abstract}
}

\emph{Introduction.}
In the past years semiconductor quantum dots have gained considerable attention 
as tunable magnetic impurities \cite{Kondo-popular}. Due to their small size electronic transport is strongly influenced by Coulomb blockade 
\cite{Coulomb-blockade}. 
A well--known  many--body phenomenon, the \emph{Kondo effect}, 
was found in quantum dots \cite{kondo} with odd 
electron number, as predicted earlier \cite{kondo-theo}. In these systems, 
a single unpaired spin is screened at low temperatures, giving rise to 
an enhanced conductance at low bias. 

Remarkably, a similar conductance enhancement was recently also observed 
for an \emph{even} number of electrons both in vertical \cite{Sasaki00} and 
lateral \cite{Schmid00} GaAs quantum dots. 
This can be understood by taking into account the  strong intra--dot electronic exchange coupling 
\cite{Tarucha00} which is ferromagnetic, similar to Hund's rule in atomic physics. 
Tuning of the level spacing by an external magnetic field then induces a singlet--triplet 
transition in the ground state. This additional degeneracy leads to the  
enhanced conductance at low temperature \cite{Eto00,Pustilnik00,Pustilnik01}.

Model calculations for the singlet--triplet transition, 
except \cite{Pustilnik01}, have so far mainly 
considered the case where the orbital quantum number characterizing 
the levels in the dot is also present in the leads. 
In particular, two conduction channels per lead have been taken into account.  
While this is appropriate for vertical 
devices \cite{Sasaki00}, recent measurements on lateral quantum dots 
\cite{vanderWiel01} suggest an interpretation in terms of a \emph{single conduction channel per lead}, 
i.e. strong orbital mixing. In the following, we will focus on this situation. 

Our main tool of analysis is Wilson's  Numerical Renormalization Group (NRG) \cite{nrg}, 
a non--perturbative approach to quantum impurity systems. In contrast to 
mean--field or scaling calculations, this method does not rely on any assumptions 
regarding the ground state or leading divergent couplings. We will find that 
this is crucial in the present analysis. 

\begin{figure}[h]
\begin{center}
\epsfig{file=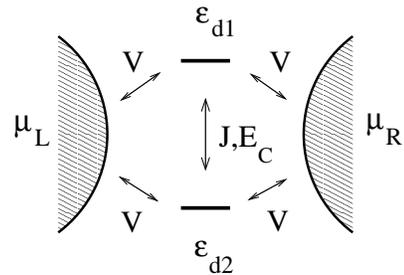,width=0.6\linewidth}
\end{center}
\caption{\label{fig1}Two--level quantum dot (\ref{dot}) with 
single--channel leads (chemical potentials $\mu_L$ and $\mu_R$). 
$V$ denotes the tunneling matrix element between the dot levels and the leads.}
\end{figure}

\emph{The Model.}
We consider a two--level Anderson impurity model as shown in fig.~\ref{fig1}. 
The Hamiltonian $H = H_L + H_R + H_D + H_T$ contains two leads 
$H_r = \sum_{k r} \epsilon_{k r} a^\dagger_{k\sigma r} a^\pdag_{k\sigma r}$ 
with $r=L/R$. The isolated dot is described by 
\beq \label{dot}
H_D = \sum_{n \sigma} \epsilon_{d n } d^\dagger_{n\sigma} d^\pdag_{n\sigma} + J \, {\bf S}_1\,  {\bf S}_2 + 
E_C \left(N - \mathcal{N}\right)^2
\eeq
where $n=1,2$ denotes the two levels, $N=\sum_{n \sigma} d^\dagger_{n\sigma} d^\pdag_{n\sigma}$ 
is the total number of electrons occupying the dot and 
${\bf S}_n = (1/2)\sum_{\sigma \sigma'} d^\dagger_{n\sigma}  
\bbox{\sigma}_{\sigma \sigma'} d^\pdag_{n\sigma'}$
are the spins of the two levels. 
Furthermore, we have introduced the \emph{charging energy} $E_C$ and an exchange 
coupling $J$ which arises due to Hund's rule. 
We choose ${\mathcal N} =2$ in order to achieve double occupancy of the dot. 
Through the energies $\epsilon_{dn}$ the level spacing $\Delta\epsilon = \epsilon_{d1} - \epsilon_{d2}$
as well as the precise position in the Coulomb blockade valley can be tuned. 
Experimentally, ${\mathcal N}$ depends on the gate voltage, while $\Delta\epsilon$ is controlled 
by an external magnetic field. 
As a consequence of Hund's rule, the intra--dot exchange is ferromagnetic ($J < 0$). 
Therefore, for $\Delta\epsilon = -J/4$ the three triplet configurations 
$|1,1> = d^\dagger_{1\uparrow} d^\dagger_{2\uparrow} |0>$, 
$|1,0> = (1/\sqrt{2}) (d^\dagger_{1\uparrow} d^\dagger_{2\downarrow} + d^\dagger_{1\downarrow} d^\dagger_{2\uparrow} |0>$, 
$|1,-1> = d^\dagger_{1\downarrow} d^\dagger_{2\downarrow}|0>$ and the singlet 
$|0,0> = d^\dagger_{2\uparrow} d^\dagger_{2\downarrow} |0>$ are degenerate. 
Motivated by the small $g$--factor in GaAs \cite{Sasaki00}, we neglect the 
Zeeman splitting of the triplet states. 
Finally, tunneling between the dot and leads is modeled by 
$H_T = \sum_{k \sigma n r} (V_{nr} a^\dagger_{k\sigma} d^\pdag_{n\sigma} + {\rm h.c.})$ where we 
neglect the energy dependence of the tunneling matrix elements $V_{nr}$ and in the following 
take them to be symmetric and identical for both levels, that is $V_{nr} = V$. 
The intrinsic line width of the dot levels due to tunnel coupling to the leads 
is $\Gamma = \Gamma_L + \Gamma_R$ with $\Gamma_{L/R} = 2\pi |V|^2 N_{L/R}$ 
where $N_{L/R}$ is the density of states in the leads. 
This model has been studied before \cite{Izumida98} without discovering the 
quantum phase transition we describe in this paper.

\emph{Transmission probability.}
We are interested in calculating electronic transport through the dot (\ref{dot}) 
close to the singlet--triplet transition. 
To this end, we use the generalized Landauer formula \cite{landauer} 
\beq \label{current}
I = \frac{2 e}{h} \int d\omega\, 
\left(f(\omega - \mu_L) - f(\omega - \mu_R)\right)\, T(\omega)
\eeq
with the Fermi function $f(\omega)$ and the transmission coefficient  
\beq \label{transmission}
T(\omega) = - \sum_{n,n',\sigma} 
\frac{\Gamma^L \Gamma^R}{\Gamma^L + \Gamma^R}\;
{\rm Im} G_{n n' \sigma}(\omega).
\eeq
Here we have introduced the retarded dot Green's functions 
$G_{n n' \sigma}(t) = -i \theta(t) <\{d^\pdag_{n\sigma}(t), d^\dagger_{n'\sigma}\}>$. 
In the following we focus on the low bias regime, where $T(\omega)$ can be 
evaluated in equilibrium, using the Numerical Renormalization Group. 
For a detailed description of this technique see Ref.~\cite{nrg}.
Note that the equilibrium transmission $T(\omega)$ also yields an approximation to the   
differential conductance $dI/dV$ at finite bias.

\begin{figure}[h]
\begin{center}
\epsfig{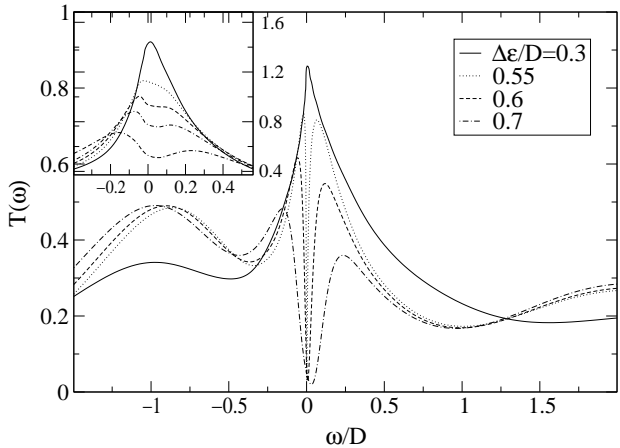}
\end{center}
\caption{\label{fig2}Transmission coefficient at zero temperature 
for different level spacings $\Delta\epsilon = \epsilon_{d1} - \epsilon_{d2}$ 
at ${\cal N}=2$, $E_C / D = 1$, $\epsilon_{d1}/D = 1$,  
$\Gamma = 0.57 D$ and $J/D = -2$. The bandwidth is given by $2 D$. 
Inset: For comparison, we show the transmission in the case of two 
conduction channels per lead, identical dot parameters, $\Gamma = 0.28 D$ 
and $\Delta\epsilon / D = 0.4, 0.5, 0.55, 0.6, 0.7$ (from top to bottom).}
\end{figure}

In Fig.~\ref{fig2} we show results for the transmission as a function of the 
level spacing $\Delta\epsilon$, corresponding to a variation of the external magnetic 
field in the experiment. Our unit is the half bandwidth $D$ of the conduction electrons. 
For $\Delta\epsilon/D \lesssim 0.5$, 
both orbitals are equally occupied and the dot is in a triplet state ($S=1$). 
Due to the hybridization with the leads, this local spin is partially screened, 
giving rise to the Kondo resonance shown in fig.~\ref{fig2} for $\Delta\epsilon / D = 0.3$. 
Note the unusual shape of this peak, which is due to the ``underscreening'' of the 
local moment -- a free spin $1/2$ remains present in the ground state \cite{Nozieres80} 
and leads to logarithmic corrections to Fermi--liquid behaviour. 
Nevertheless, as for the spin $1/2$ Kondo effect, the transmission reaches the 
unitary limit at low temperatures. Due to systematic numerical errors 
in the NRG calculation, this limit is underestimated by about $10\%$. 

In the regime $\Delta\epsilon/D \gtrsim 0.5$, both electrons occupy the lower dot level 
and the ground state is a singlet. Remarkably, in this case a \emph{pronounced dip} 
arises \emph{within the Kondo peak}, leading to a strongly reduced transmission at 
low energy. The residual value $T(0)$ is independent of $\Delta\epsilon$ 
and is determined by the position in the Coulomb blockade valley. 
In particular, it vanishes in the center of the valley 
where $\epsilon_{d1} = -\epsilon_{d2}$.  

In order to demonstrate clearly the importance of the 
number of conductance channels, we show results for the case of two 
transmission channels onto the dot (\ref{dot}) in the inset of fig.~\ref{fig2}.
In this case, the local spin is always completely screened by the leads. 
One obtains a conventional Zeeman type splitting of the conductance peak due to the 
energy difference between singlet and triplet. The sharp dip described above 
does not occur here and is thus \emph{characteristic for the single channel situation}. 
\begin{figure}[h]
\begin{center}
\epsfig{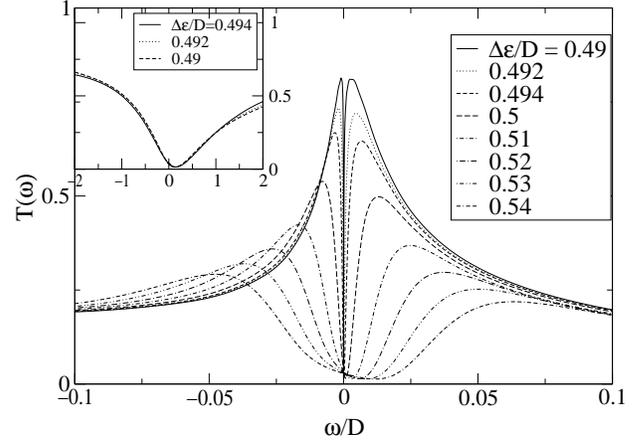}
\end{center}
\caption{\label{fig3}Scaling of the dip at zero temperature. 
Parameters are chosen as in fig.~\ref{fig2}, but with a smaller broadening 
$\Gamma = 0.25 D$. Note the ``pinning'' of the transmission at $\omega=0$. 
Inset: Rescaled dip $T(\omega/\omega^*)$ 
for parameter values close to the transition. 
The curves collapse onto a universal scaling function with a single parameter 
$\omega^*$ (dip width).
}
\end{figure}
We find that for small level spacing the dip has a \emph{universal scaling form} 
which is completely determined by its width $\omega^*$.  This can 
be seen more clearly in fig.~\ref{fig3}, where we focus on the singlet side 
close to the transition. The scaling curve is extracted in the inset. 

\begin{figure}[h]
\begin{center}
\epsfig{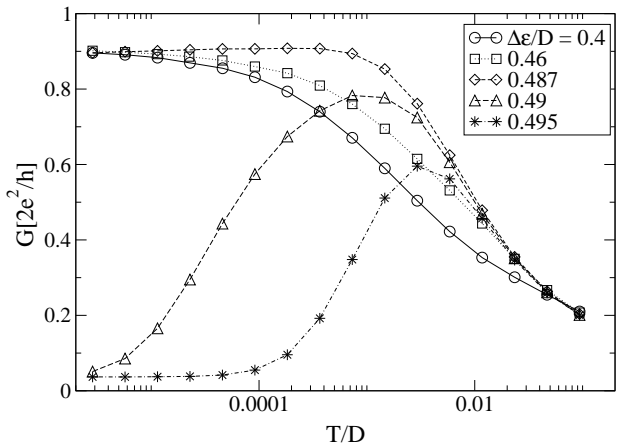}
\end{center}
\caption{\label{fig4}Linear conductance for identical parameters as in fig.~\ref{fig3} 
in three different regimes: on the triplet side ($\Delta\epsilon/D = 0.4, 0.46$), at 
the transition $(\Delta\epsilon/D = 0.487$) and on the singlet side 
($\Delta\epsilon/D = 0.49, 0.495$).}
\end{figure}

\emph{Linear conductance.} 
Using the current formula (\ref{current}), we now determine the behaviour 
of the conductance as a function of temperature. Results are shown in fig.~\ref{fig4}. 
In the triplet regime, upon lowering of the temperature, 
the conductance rises monotonously up to the unitary limit 
due to the partial screening of the local spin $S=1$. 
Note that the associated Kondo temperature $T_K$ is extremely low. This may be the 
reason why the triplet Kondo effect has so far not been observed experimentally, 
though some indications have been seen in Ref.~\cite{Sasaki00}. 
Close to the singlet--triplet transition, $T_K$ is strongly enhanced. 
On the singlet side, we find a ``bump'' type behaviour of the conductance when $T$ is lowered: 
After an initial rise due to the Kondo effect, $G(T)$ decreases strongly at $T \lesssim T^*$. 
with a small residual value for $T\to 0$ determined by the position in the Coulomb blockade valley.  
Note that, like the increase, the decrease of $G(T)$ is logarithmic, indicating a 
\emph{two--stage Kondo effect}. 
In particular, the $T\to 0$ behaviour of $G(T)$ is again universal and can be characterized 
by a single fit parameter $T^* \sim \omega^*$. 

\emph{Quantum phase transition.} 
Here we present a physical explanation of the above results. It has been suggested 
earlier \cite{Kikoin01} that the singlet--triplet degeneracy of the dot 
can be parametrized in terms of a  two--spin $S=1/2$ Kondo model. 
Formally, this is achieved by performing a Schrieffer--Wolff projection \cite{Schrieffer66} 
of our two--level Anderson Hamiltonian on the (almost) degenerate subspace spanned by the four 
states $|0,0>$, $|1,1>$, $|1,0>$ and $|1,-1>$. 
One obtains an effective Hamiltonian of the following form 
\beq \label{Kondo}
H_{\rm eff} = H_{L} + H_{R} + J_1 {\bf \tilde{S}}_1 \bbox{s} + J_2 {\bf \tilde{S}}_2 \bbox{s} + 
I {\bf \tilde{S}}_1 {\bf \tilde{S}}_2.
\eeq
where both Kondo spins are coupled to the same conduction channel 
$\bbox{s} \equiv (1/2N)\sum_{k k'} a^\dagger_{k\sigma} \bbox{\sigma}_{\sigma \sigma'} a_{k'\sigma} $.
Here, $N$ is the number of $k$-states in the leads and 
we have already taken a symmetric combination of left and right lead 
according to $a_{k\sigma} = (a_{k\sigma L} + a_{k \sigma R})/\sqrt{2}$. 
Additional potential scattering terms have been neglected. 
Note that the ${\bf \tilde{S}}_1$, ${\bf \tilde{S}}_2$ introduced above are fictitious spins, 
different from the original levels. 

To leading order, the parameters in (\ref{Kondo}) are given by the following expressions
(note that $V \sim 1/\sqrt{N}$):
\bea
J_{1(2)} &=& 
2 N V^2 \Big(\frac{1 \mp \sqrt{2}}{\epsilon_{d1} + E_C - J/4} + 
\frac{1}{\epsilon_{d2} + E_C - J/4} \\
&& - \frac{1}{\epsilon_{d1} - E_C + J/4} - 
\frac{1 \mp \sqrt{2}}{\epsilon_{d2} - E_C + J/4} \Big) \nonumber
\eea
\bea
I &=& 2 N V^2 \Big(\frac{1}{\epsilon_{d1} - E_C + J/4} + 
\frac{1}{\epsilon_{d2} - E_C + J/4}  \nonumber \\
&& - \frac{2}{\epsilon_{d2} - E_C}\Big) + \Delta\epsilon + \frac{J}{4}.
\eea
In particular, we find $J_1 \ne J_2$. The effective direct exchange $I$ 
will be a function of the level splitting $\Delta\epsilon$ in the original model.

The Hamiltonian (\ref{Kondo}) has been analyzed recently \cite{two-impurity-kondo}. 
Depending on the strength of $I$, 
the ground state of the two spins is either an inter--impurity--singlet or a triplet. 
The associated transition at $I = I_{\rm crit}$  is of the \emph{Kosterlitz--Thouless} type. 
The triplet side corresponds to an underscreened $S=1$ Kondo model, while on the 
singlet side, a \emph{two--stage screening} process of the two impurities has been found 
for small $\Delta I \equiv I - I_{crit} > 0$. 
First, the larger one of the two couplings (e.g. $J_1$) leads to a screening of the 
corresponding spin ${\bf \tilde{S}}_1$ by the Kondo effect, thus decoupling ${\bf \tilde{S}}_2$ 
from the conduction band for $T < T_K$.  
The effective inter--impurity exchange $\Delta I$ 
then leads to a second Kondo effect for ${\bf \tilde{S}}_2$. 
At low temperatures, the two spins form a singlet with a binding energy 
\beq
T^*\sim \exp(-T_K/\Delta I) 
\eeq
that is indeed exponentially small in the distance 
$\Delta I \approx \Delta\epsilon - \Delta\epsilon_{\rm crit}$ from 
the critical point. 
This argument holds only as long as $\Delta I < T_K$. For larger values of the 
effective exchange, $\Delta I$ provides a cutoff on the ``first'' Kondo effect 
and the singlet binding energy is then linear, $T^* \approx \Delta I$, as a function 
of the level splitting. 
In both cases, transport at $T < T^*$ is strongly suppressed due to the singlet formation 
which leads to the dip in the density of states.

\begin{figure}[h]
\begin{center}
\epsfig{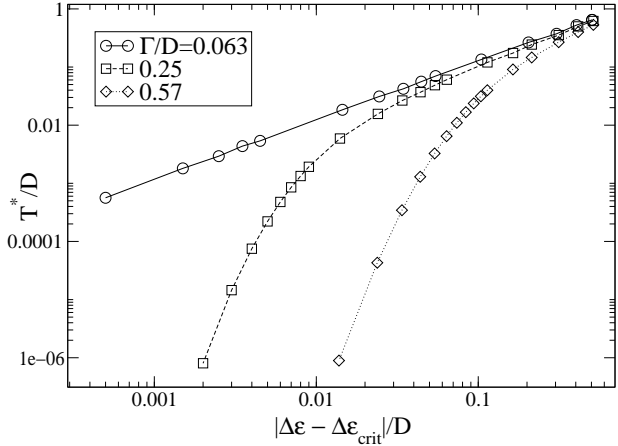}
\end{center}
\caption{\label{fig5}Scaling of the characteristic temperature scale $T^*$ 
(width of the dip) versus the distance from the transition point 
for $E_C/D = 1$, $\epsilon_{d1}/D = 1$ and $J/D = -2$. }
\end{figure}

In order to demonstrate that this is the correct low--temperature scenario  
of the two--level quantum dot, we have calculated the characteristic 
energy scale $T^*$ determined by the width of the dip. 
This calculation has been performed for the full dot Hamiltonian (\ref{dot}).
In fig.~\ref{fig5}, we give $T^*$ as a function of the distance from the critical point.
Clearly, for large $|\Delta \epsilon - \Delta \epsilon_{\rm crit}|$ the dip scales 
linearly, while close to the critical point a crossover to an exponential dependence 
occurs. Note that different level broadenings lead to largely different Kondo temperatures; 
in particular, for $\Gamma/D = 0.063$ only the linear behaviour of $T^*$ is seen 
because $T_K$ is extremely small.

At this point we would like to point out the robustness of our results 
with respect to parameter changes in the model. We have 
chosen our quantum dot to be at a generic position in the Coulomb 
blockade valley, thus demonstrating that the quantum phase transition 
and the suppression of low--temperature transport discussed here 
are not restricted to special situations like particle--hole symmetry. 
The dip is also found when the broadening of the two levels is tuned 
to different values and/or when an asymmetry between the coupling to the right and left 
lead is introduced. 

\emph{Conclusion and experimental relevance.} 
Motivated by recent transport experiments \cite{vanderWiel01} 
we have studied transport through a lateral quantum dot 
modelled by two levels close to the Fermi surface coupled 
to a single conduction channel in the leads. Correlations 
between different electrons in the dot are taken into account via the 
charging energy and a ferromagnetic exchange coupling due to Hund's rule. 

At the associated singlet--triplet degeneracy, we find a 
\emph{quantum phase transition} of the Kosterlitz--Thouless type, 
as in the two--impurity model \cite{two-impurity-kondo}. 
On the triplet side of the transition the conductance 
simply increases up to the unitary limit upon lowering of $T$.
On the singlet side, we find  a non--monotonic behaviour of the conductance 
as a function of temperature 
(``bump'') corresponding to a characteristic ``dip'' in the 
transmission, which is also expected to be seen in the differential conductance. 
In particular, the width of the dip represents a new low--energy scale 
-- the singlet binding energy -- which becomes exponentially small 
close to the transition. For two conduction channels, none of the two 
effects is observed. 

These findings are in good agreement with recent conductance measurements 
for a lateral quantum dot \cite{vanderWiel01} close to the singlet--triplet 
transition. In this system, orbital symmetry is not conserved during tunneling. 
Both the non--monotonic behaviour of $G(T)$ and the sharp dip in $dI/dV$ have been observed, 
in contrast to previous studies of vertical quantum dots \cite{Sasaki00}. 

We therefore conclude that the number of conduction channels plays a crucial role  
for low--energy transport properties of a quantum dot. 
Symmetry and physical behaviour of such a device are thus strongly related.

\emph{Acknowledgements.}
The authors would like to thank W.~van der Wiel, S.~De Franceschi, and 
M.~Vojta for useful discussions. This work is supported by the 
Deutsche Forschungsgemeinschaft under SFB 484 (W.H.).

\end{document}